\documentclass[aps,notitlepage,twocolumn,nofootinbib,superscriptaddress,pra,10pt]{revtex4-1}
\usepackage{amsmath,amsfonts,amssymb,amsthm,bbm,graphicx,enumerate,times}
\usepackage{mathtools}
\usepackage[usenames,dvipsnames]{color}

\usepackage{hyperref}
\hypersetup{
	breaklinks,
	colorlinks,
	linkcolor=gray,
	citecolor=gray,
	urlcolor=gray,
}
\usepackage{tikz}
\usepackage{graphicx}
\usepackage{float}
\usepackage{comment}
\usepackage{csquotes}
\usetikzlibrary{shapes,positioning}
\usepackage{etoolbox}
\definecolor{cadmiumgreen}{rgb}{0.0, 0.42, 0.24}
\definecolor{nick}{rgb}{0.8,0.4,0.1}

\newcommand{\co}{CO$_2$}

\include{math_defs}
\newcommand{\id}{\mathbb{I}}

\newcommand{\ket}[1]{|#1\rangle}
\newcommand{\bra}[1]{\langle#1|}
\newcommand{\ketbra}[2]{|#1\rangle\langle#2|}

\newcommand{\tr}{\text{ tr }}

\newcommand{\fr}{\mathcal{F}}

\graphicspath{{./fig/}}

\begin{document}
\title{Harnessing symmetry-protected topological order for quantum memories}
\author{M. Goihl}

\affiliation{Dahlem Center for Complex Quantum Systems, 
Freie Universit{\"a}t Berlin, 14195 Berlin, Germany}

\author{N. Walk}

\affiliation{Dahlem Center for Complex Quantum Systems, 
Freie Universit{\"a}t Berlin, 14195 Berlin, Germany}

\author{J. Eisert} 

\affiliation{Dahlem Center for Complex Quantum Systems, 
Freie Universit{\"a}t Berlin, 14195 Berlin, Germany}

\affiliation{Helmholtz-Zentrum Berlin f{\"u}r Materialien und Energie, 14109 Berlin, Germany}

\author{N. Tarantino}

\affiliation{Dahlem Center for Complex Quantum Systems, 
Freie Universit{\"a}t Berlin, 14195 Berlin, Germany}

\date{\today}

\begin{abstract}
Spin chains with symmetry-protected edge modes are promising candidates to
realize intrinsically robust physical qubits that can be used for the storage and processing of
quantum information. In any experimental realization of such physical systems,
weak perturbations in the form of induced interactions and disorder are
unavoidable and can be detrimental to the stored information. 
At the same time, the latter may
in fact be beneficial; for instance by deliberately inducing disorder which
causes the system to localize. In this work, we explore the potential of using an 
$XZX$ cluster Hamiltonian to encode quantum information into the local edge 
modes and comprehensively investigate the influence of both many-body interactions 
and disorder on their stability over time, adding substance to the narrative
that many-body localization may stabilize quantum information. 
We recover 
%tomographic 
the edge state at each time step, allowing us to reconstruct the quantum channel that captures the
locally constrained out of equilibrium time evolution.
With this representation in hand, we analyze how well classical and quantum
information are preserved over time as a function of disorder
and interactions. We find that the performance of the edge qubits varies dramatically between disorder realizations. Whereas some show a smooth decoherence over time, a sizeable fraction are rapidly rendered unusable as memories. We also find that the stability of the classical information -- a precursor for the 
usefulness of the chain as a quantum memory -- depends strongly on
the direction in which the bit is encoded. When employing the chain as a genuine 
quantum memory, encoded qubits 
are most faithfully recovered for low interaction and high disorder.
\end{abstract}
\maketitle

\begin{figure}
\includegraphics[width=0.45\textwidth]{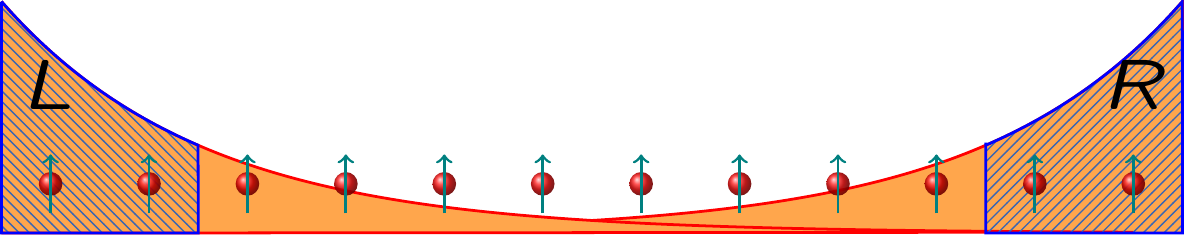}
  \caption{
  This is a sketch of the one-dimensional spin chain investigated in this work.
  The support of the edge zero mode operators under the influence of perturbations 
  is depicted in orange. The blue
  regions are the local edge  qubits to which an experimenter would have access.
}
\label{f:setup}
\end{figure}

\section{Introduction}
\label{sec:intro}

The prospect of highly controllable quantum simulators \cite{Gross:2017do,Blatt:2012gw,AspuruGuzik:2012jm,1408.5148} allows for the simulation of complex Hamiltonians, potentially realizing exotic phases of
matter. A celebrated class of such phases are topological phases of matter which offer the promise of robust encodings of quantum information with a reduced need for active error correction \cite{TQMemory2002,Kitaev-AnnPhys-2003,kitaev2006topological}. 
Simple instances of these exist in one-dimensional spin chains, where one finds 
\emph{symmetry-protected topological phases (SPTs)} 
which host emergent edge zero modes that enforce approximate degeneracies in the spectrum and define edge states which are dynamically decoupled from the bulk \cite{haldane1983nonlinear,kitaev2001unpaired,PhysRevB.96.165124,Wen_Anomalies_2013,PhysRevB.87.155114}. The latter feature makes these systems interesting 
candidates for physical qubits used in quantum memories, as well as in implementations of measurement-based 
quantum computation \cite{Miller_Miyake2015,Stephen_Wang_Prakash2017,Raussendorf_Wang_Prakash2017}, which exploit the entanglement 
properties of the SPT to effectuate gate operations. However, for a practical quantum memory architecture, the topologically facilitated isolation of the edge states from the bulk must preserve 
the encoded data for a sufficiently long time, ideally several thousand multiples of 
the time taken to implement logical gates.

The ability of an SPT edge mode to retain information depends largely on the strength of interactions which perturb the SPT order. These drive two related phenomena: the splitting of the ground state degeneracy and the broadening of its 
\emph{edge zero modes}.   
While the former controls the phase transition, the latter can have a significant impact on our ability to practically use the edge modes for storage purposes. We should typically expect to encode information into only a finite number of spins located at the boundary of the chain. 
The edge mode has no such restriction, and usually decays exponentially into the chain (see Fig.\,\ref{f:setup}).
This mismatch between the finite number of sites we can encode into and the long tails of the edge mode operators means 
that we can never expect our information to be totally preserved even when we take the system to the thermodynamic limit to make the edge modes exact constants of motion. With this in mind, one naturally turns to search for features which can further localize the edge modes without engaging in fine-tuning.

One readily available and promising candidate for this role is \emph{disorder}. While usually considered an obstacle to 
analytical and numerical study, this feature is both generically present in physical systems and 
also enables a variety of localization phenomena, which may conceivably
enhance the localization of the emergent edge modes. These go by different names depending on 
whether the underlying physics is describable using free fermions (\emph{Anderson
localization} \cite{Anderson1958}) or 
is intrinsically interacting (\emph{many-body localization, abbreviated to MBL} \cite{Abanin2019,christian_review}). 
While the exact phenomenology of these two differ, both of those feature local constants of motion.
In the SPT context this has been proposed as a method to attenuate the effective interaction 
between the two edge modes, since any coupling between them must be mediated by bulk eigenstates
which, if only locally supported, can only connect the two via a prohibitively
small number of virtual processes \cite{PhysRevB.88.014206,Chandran14,Kjaell14}.

While this heuristic is appealing, precious little confirmation of it has been forthcoming 
since initial investigations established that the edge modes are not de-stabilized 
by the presence of disorder \cite{bahri2015localization}. In fact, a recent
work by some of the authors \cite{ourSPT} -- built on the earlier 
Ref.~\cite{GoihlNew} -- suggests that the shape of the edge mode
when subjected to disorder and interaction changes in a more intricate way
than expected . There, the support of the edge modes was
investigated and, surprisingly, found to depend heavily on the specific edge mode 
that was analysed.
In light of this, a thorough 
examination of disorder effects on the temporal stability of edge mode appears to be
urgently required. Here, achieve this goal: We explicitly determine the dynamics of the encoded subspace 
and quantify the extent to which information can
be recovered at a given time.

Before proceeding, it is critical to specify what is meant by a good quantum memory. 
Any imperfect memory device can be viewed as a partially decohering channel acting upon the 
encoded quantum states. For a comprehensive investigation, the first step is to compute the operators 
that define this channel, from which all potentially relevant measures can then be calculated. 
It is important to carefully select measures that meaningfully quantify the notion of information 
preservation in the context of quantum computation and communication applications. 
Ideally, a measure should be operationally well motivated, straightforward to calculate and amenable 
to experimental verification. In this work, we perform full
reconstruction of the time-evolved state on the edge mode subspace. This
allows us to fully reconstruct the effective quantum channel applied at each time step. 
With this reconstruction at hand, we compute bounds on two measures of information preservation 
when using the system as a quantum memory: the retrievability of a \emph{logical
bit}, which is a prerequisite for a quantum memory; and the \emph{coherent information}, which bounds the 
\emph{quantum capacity}, that is the ultimate limit to quantum information transmission and particularly 
relevant for quantum communication applications, e.g., 
quantum repeaters \cite{Briegel:1998jd,Muralidharan:2016ky}. For the classical information, 
we find a multi-faceted behaviour consistent
with Ref.\,\cite{ourSPT} supporting the observation that encoding with
different edge state bases results in different stability behaviour. In the quantum
information setting, we find that disorder helps the recovery of the encoded
qubits, but low interaction strength is mandatory.
Our results extend and enrich the narrative of 
disorder helping localize edge modes, as we find that there are 
encodings and Hamiltonian parameters for which this paradigm holds 
and others where it does not. In practice, this either makes an optimization
over all possible parameters necessary to find the \enquote{sweet spot}
of information coherence or demands an educated choice of physical realization. 
One can equally see our work as shedding new lights into notions of locally
constrained non-equilibrium dynamics that interplays with disorder \cite{PhysRevLett.121.085701}.

\section{Setting}

In this work, we analyze the interplay of disorder, interactions and symmetry-protected topological (SPT) order.
We are specifically interested in the stability 
of the protected subspace spanned by the unperturbed edge modes when the system is subjected to many-body interactions and disorder. 
To measure the extent to which the locally 
encoded state gets entangled with the bulk, we time evolve the system 
and reconstruct the unperturbed edge mode subspace.
We study a spin chain of length $N$ hosting a disordered $XZX$ cluster Hamiltonian 
\begin{equation} 
H_{0}(\{h_j\})= - \sum_{j=2}^{N-1} (1+h_j) X_{j-1}Z_j X_{j+1}\,,
\label{eq:HXZX} 
\end{equation} 
where the $h_j$ are drawn uniformly from the interval 
$[-\frac{\Delta}{2},\frac{\Delta}{2}]$ and
$X_j, Y_j, Z_j$ are the Pauli operators acting at site $j$. 
Here, we restrict the disorder to act only on the cluster terms, disordering around a mean bulk gap of 1. We include this offset to distinguish the ability of disorder to help localize the edge modes  from the bulk energy gap's power to do the same.
The choice to have disorder appear in the coefficient of the cluster terms is a pragmatic one; any other model of disorder, e.g. a random local magnetic field, competes 
with the SPT order and thereby drives a transition to a topologically trivial phase 
\cite{PhysRevB.88.014206,bahri2015localization}. 
Disordering the cluster terms themselves is an ideal version of disorder in the
sense that this version splits degenerate levels for excited states, while
preserving the ground state manifold. 
Therefore, we expect any other model of disorder to cause less localization in
the system.

Even disordered, the Hamiltonian in Eq.\,\eqref{eq:HXZX} is a solvable 
representative of an SPT phase protected by the time reversal operator 
\begin{equation}  
 \mathcal{T} = \prod_{i=1}^N Z_i \mathcal{K}\,, \label{eq:TR} 
\end{equation} 
where $\mathcal{K}$ is complex conjugation. Thus, this system supports 
protected edge zero modes, operators localized on the edge which commute with the 
Hamiltonian, at least up to an error that is exponentially suppressed in system
size. 
At this solvable point, these operators commute exactly and can be 
identified by inspection. There are six such operators, 
\begin{align} 
\mathcal{O}_{\rm Edge}^i = \left\{ \begin{array}{ cccc} \mathcal{O}_{L}^x = X_1, &\mathcal{O}_{R}^x = X_N, \\
 \mathcal{O}_{L}^y = Y_1 X_2, &\mathcal{O}_{R}^y = X_{N-1}Y_N ,\\ 
\mathcal{O}_{L}^z = Z_1 X_2, &\mathcal{O}_{R}^z= X_{N-1}Z_N 
\end{array} \right\}\,, 
  \label{e:qb}
\end{align}  
that split into two independent Pauli algebras, one for the left and right edge, 
respectively. Each of these Pauli algebras implies the existence of a special 
two-dimensional Hilbert space hosted on these edges, corresponding to the fundamental 
representation of these algebras.

The physical situation underlying this algebraic picture 
is reflected by a subtle decomposition of the Hilbert space. 
The \emph{logical} left (right) edge qubit is only encoded into a two-dimensional subspace 
${\cal H}_{ L}$ (${\cal H}_{R}$) of the four-dimensional Hilbert space described 
by the physical qubits on the two left-most (right-most) sites. It is important to stress
that the bulk degrees of freedom, associated with 
${\cal H}_\text{Bulk}$, involve physical sites $3,\dots, N-2$ \emph{as well as} 
the remaining two-dimensional subspaces of each of the two edges.
Initial preparations are  product state vectors of the form
\begin{equation}
\label{preparation}
  |\psi\rangle = |\psi_{L}\rangle \otimes |\psi_\text{Bulk}\rangle  \otimes
  |\psi_{R}\rangle\,.
\end{equation}
In the ideal limit, without any interactions,
these edge states $|\psi_{L}\rangle$, $|\psi_{R}\rangle$ are independent of 
the global unitary time evolution generated by $H_0$ -  never becoming
entangled with the bulk -  and thus do not decohere. 
This changes, however, as soon as coupling terms are added to the Hamiltonian. 
In this case, the actual edge modes 
broaden (sketched in Fig.\,\ref{f:setup}) and no longer coincide with their 
compact form in Eq.\,\eqref{e:qb}. The corresponding edge states hybridize, 
splitting the degeneracy by an amount which is suppressed exponentially 
with system size
\cite{haldane1983nonlinear,kitaev2001unpaired,affleck1988valence,affleck1987rigorous}.
Our logical qubits are no longer time independent and become entangled with the 
rest of the chain. By restricting ourselves to measurements of the edge operators 
in Eq.\,\eqref{e:qb}, we will be partially tracing an entangled state
leading to decoherence over time.

To model this effect, we introduce an Ising-type interaction 
\begin{equation} 
H_\text{ZZ}(J)=- J \sum_{j=1}^{N-1} Z_j Z_{j+1}\,, 
\label{eq:HZZ} 
\end{equation}
resulting in the following Hamiltonian 
\begin{equation} 
H(\{h_j\},J)=H_0(\{h_j\})+H_\text{ZZ}(J)\,,
\label{eq:H} 
\end{equation} 
where we vary both $J$ and $\Delta$. Note that the resulting Hamiltonian
is not free anymore and hence requires a treatment in the full Hilbert space.

As we are interested in studying the evolution of the states of the edge 
Hilbert space, we need a way of constructing the reduced dynamics on what 
is effectively an open quantum system. This can be done by time evolving an
initial pure product state on the entire system of the form in Eq.\,\eqref{preparation}, 
and then evaluating the time-evolved reduced state of the edge, $\rho_t$, as
\begin{align}
\label{e:chan}
\rho_t &= \tr_\text{Bulk} (e^{-iHt} \ket{\psi} \bra{\psi} e^{iHt} ) \\
&= \mathcal{E}_t(|\psi_{L,R}\rangle \langle \psi_{L,R}|)\,, \nonumber
\end{align}
with $|\psi_{L, R}\rangle=|\psi_{L}\rangle  \otimes |\psi_{R}\rangle$.
%as a time evolving state $\rho_t$ defined on the Hilbert space ${\cal H}_{L,R}={\cal H}_L\otimes {\cal H}_R$
%of the unperturbed logical edge modes given by Eq.\,\ref{e:qb}.
These dynamics, for fixed bulk preparations, gives rise to 
a family of quantum channels ${\cal E}_t$, acting on
the state space over ${\cal H}_{L,R} $. This time-parametrized 
family is also referred to as the \emph{dynamical map} 
that captures time evolution of the edge. 
We now investigate the effects of these dynamics on the viability of the edge modes as a quantum memory.
%We now turn to the question of what this evolution means for information stored 

\section{Methods}

\subsection{Characterizing quantum memories} \label{ssec:QMem}
The approximate conservation of the observable quantities 
is often used as a proxy for the information 
storage capabilities of a physical system, topological or otherwise. In fact, disordered
models in a many-body localized phase have been shown  
to preserve spin orientation along a magnetization direction both theoretically 
\cite{Serbyn2013,Huse2014} and experimentally \cite{Schreiber2015} in one 
dimension, with similar results in two dimensions \cite{Choi2016}. While these phenomena are 
indeed hallmarks of dynamical decoupling,
they are only a prerequisite for a useful quantum memory. Specifically,
the information encoded into a polarized spin state in an MBL system
is only a classical bit string, even if the system itself is quantum. Given the 
abundance of efficient classical memories, using a quantum system for bit 
storage seems wasteful. A true quantum memory must be able to 
robustly store an arbitrary - typically unknown - qubit. This necessitates 
the existence of a protected two-dimensional Hilbert space, which is not 
guaranteed by the presence of even an arbitrary number of conserved quantities. Rather, preservation of any fixed basis can be seen as a necessary condition,
but to promote the system in question to a quantum memory, all other
possible bases must be preserved, too. 

To assess the usefulness of a system as a 
quantum memory, we will now introduce measures which quantify the amount of
both classical and quantum information
that are retrievable
from our system. As mentioned above, classical information storage is a necessary
condition for realizing a quantum memory.
The storage capabilities of quantum states are relevant for 
quantum computations, where one reads in a classical bit string into a 
quantum system, then performs some
(possibly) quantum gates on it and reads out the classical bit string encoding the
result of the computation again. 
Moreover, many quantum communication protocols
require a quantum memory, e.\,g.\;for quantum networks it is often necessary 
to store an entangled state between some nodes while preparing further entanglement between others.
As such, quantum memories are a crucial building block
for many quantum technologies \cite{Roadmap}.

Since we are interested in the stability of information (whether classical or quantum) stored in edge states, care should be taken in defining measures thereof.
The robustness of a logical bit, whose value is encoded
via orthogonal state vectors $\ket{\psi}$ and $\ket{\phi}$ in an imperfect quantum memory described 
in its time evolution by 
$\mathcal{E}_t$, is essentially given by the distinguishability of the logical states at the 
output (i.e. between $\rho = \mathcal{E}_t(\ket{\psi}\bra{\psi})$ and 
$\sigma = \mathcal{E}_t(\ket{\phi}\bra{\phi})$. This can be
captured by the trace distance,
\begin{equation}
\mathcal{D}(\rho,\sigma) = \frac{1}{2} ||\rho-\sigma||_1\,.
\end{equation}
Operationally, this is directly related to the probability of successfully recovering an encoded logical bit.
For example, the maximum probability $p$ of distinguishing between quantum states $\rho$ and $\sigma$
on a single-shot level is
\begin{equation}
	p= \frac{1}{2}  + \frac{1}{2} \mathcal{D}(\rho,\sigma)\,,
	\label{pdist}
\end{equation}
assuming each state occurs with initial identical probability of $1/2$. This, then, is precisely the optimal probability that an unknown equiprobable logical bit could be successfully determined from the output of the memory. 

As outlined above, these notions can be extended to quantify the performance of a quantum memory by considering all possible encodings. This insight was used in Ref. \cite{Xu_2018} to propose a figure of merit called the \emph{integrity}, which is essentially the output distinguishability minimized over all possible orthogonal state encodings. Intuitively, this
measure tracks how well the entire logical subspace is preserved under the time
evolution. It can also be related to other important metrics in quantum information processing such as the logical fidelity and the error correction pseudo-threshold \cite{Svore:2005ut,Cross:2009wv}.

For our system the integrity at time $t$ will be given by,
\begin{equation}
  \mathcal{I}(t) = \min_{|\psi_{L,R} \rangle \perp
 |\phi_{L,R} \rangle}
 \mathcal{D}(\mathcal{E}_t( |\psi_{L,R}\rangle\langle \psi_{L,R}|),
   \mathcal{E}_t( |\phi_{L,R}\rangle\langle \phi_{L,R}|))\,.\label{integrity}
\end{equation}
where $ |\psi_{L,R}\rangle $
is a quantum state vector of ${\cal H}_{L,R}$ 
and $ |\phi_{L,R}\rangle $ is an orthogonal state vector, evolved in time. Note that this is a slight generalization with respect to 
\cite{Xu_2018} where only single qubit systems were considered, for which the 
orthogonal complement of a pure state is uniquely defined. The generalization 
here to two qubit states will thus require simultaneous
minimization over pairs of orthogonal input states, which can still be captured as an efficiently
solvable semi-definite
problem. Nevertheless, the primary minimization in the integrity is highly non-trivial for general $\mathcal{E}_t$ so we will resort to upper bounding this quantity by evaluating the output distinguishability for particular orthogonal encodings.

Given the above definition of the integrity, there is an immediate connection to the
\emph{classical information capacity} of the quantum channel ${\cal E}_t$. This quantifies
the asymptotic rate at which classical bits encoded in quantum states can be retrieved,
once the quantum systems undergo an evolution dictated by the quantum channel ${\cal E}_t$.
This is the most meaningful definition of capturing the capability of storing classical
information. On the formal level, it is defined as the \emph{regularized Holevo-$\chi$} of the quantum
channel \cite{MarkWilde}. Since the integrity gives rise to a specific product 
encoding, we have
%\begin{equation}
%	{\cal I}(t)\leq C_{\rm Cl}(\mathcal{E}_t)\,.
%\end{equation}
\begin{equation}
	p_{\mathcal{I}}(t)\log p_{\mathcal{I}}(t) + (1-p_{\mathcal{I}}(t))\log(1-p_{\mathcal{I}}(t))\leq C_{\rm Cl}(\mathcal{E}_t)
\end{equation}
where $p_{\mathcal{I}}(t)$ is the distinguishing probability associated with $\mathcal{I}(t)$ via Eq.~\ref{pdist}.

In what follows, we restrict ourselves to eigenstates of the unperturbed edge
mode operators in Eq.\,\eqref{e:qb} as inputs. That is, the initial orthogonal state vectors $|\psi_{L,R}\rangle$ and $|\phi_{L,R}\rangle$
entering the definition of the integrity are chosen to be
orthogonal eigenvectors of $X_L \otimes X_R$,
$Y_L \otimes Y_R$, and
$Z_L \otimes Z_R$. We make this choice because it should be 
relevant in the context of possible experimental realizations, and also because similar input states have been shown to saturate the minimum in 
Eq.\,\eqref{integrity} for some classes of channels \cite{Xu_2018}. We refer to these physically motivated variants as \emph{directed} integrities and will denote them as ${\cal I}_X(t),{\cal I}_Y(t),{\cal I}_Z(t)$, respectively.

For quantum communication, arguably the most meaningful quantity is the \emph{quantum channel capacity $Q$}, which is the maximum number of qubits which can be reliably transmitted per use of the channel in the asymptotic limit of many uses. More precisely this means that $Q(\mathcal{E}_t)$ is the maximum rate at which one can send qubits through a channel $\mathcal{E}$, such that the probability of error goes to zero in the limit of many uses of the channel. To date, this quantity has only be evaluated for certain classes of channels. This is because, in general, it requires a complicated optimization over input encodings (possibly entangled across channel uses). Typically, one writes $Q$ using a `regularization' of the the single-shot capacity $Q_1$
\begin{equation}
Q(\mathcal{E}_t) = \lim_{n\rightarrow\infty} \frac{1}{n} Q_1(\mathcal{E}_t^{\otimes n})
\end{equation}
where $\mathcal{E}_t^{\otimes n}$ represents $n$ parallel uses of the channel.
To calculate $Q_1$ one only optimizes over encodings for a single input round thus,
\begin{eqnarray}
  Q_1(\mathcal{E}_t) &=& \sup_\rho \left(S(\mathcal{E}_t(\rho))-
  S((\mathcal{E}_t\otimes \id )(\ketbra{\psi}{\psi}))\right)\nonumber \\
  &=& \sup _\rho C(\mathcal{E}_t)_{\rho}\,,
\end{eqnarray}
where $\ket{\psi}$ is a purification of $\rho$ and $S(\rho) = -\mathrm{tr}(\rho\log\rho)$ is the von Neumann entropy. The quantity on the right, $C(\mathcal{E}_t)_\rho$, is called the coherent information \cite{MarkWilde}. 

Even this optimization can be quite challenging, so we instead choose a particular input state which yields a lower bound to the channel capacity. We choose to input one half of a maximally entangled state, such that the state going through the channel, $\rho$, will be the maximally mixed state. This state is completely isotropic and should hence be a good candidate to assess the (possibly asymmetric) induced noise. In fact, it also can be shown via the Choi-Jamio\l{}kowski isomorphism that the resulting bipartite output state is a unique alternative representation of the map $\mathcal{E}$. In any event, this procedure yields a convenient lower bound to the quantum capacity of the memory.  

%\cite{nielsen} and will be denoted as $C(t)$. }
%\je{[Will do; I will also clean up this section a bit
%and give perspective what it all means.]}

\subsection{Computing figures of merit}

Having identified meaningful quantities for determining whether information is
protected, we turn to the messy business of actually computing them. The first
step is to compute the quantum channel $\mathcal{E}_t$ which captures the edge
state dynamics. There are several mathematically equivalent ways to represent a
quantum channel, such its Choi-Jamio\l{}kowski isomorphic quantum state or its
Kraus operators \cite{nielsen}. We will employ an alternative approach,
better-suited to the particularities of problem, known as the \emph{matrix
form} of the channel. Obtaining this representation will essentially boil down
to computing the final edge states given a set of well chosen initializations.
While this is formally done by evaluating the partial trace in Eq.\,\eqref{e:chan}, 
doing so directly is very inefficient. The difficulty comes from attempting to trace over ${\cal H}_\text{Bulk}$, which contains information that lives on the same \emph{physical} sites as the edge qubits. This means that we cannot use the physical site index to perform the partial trace, forcing us to change to a basis where the decomposition of our full Hilbert space into ${\cal H}_\text{Bulk} \otimes {\cal H}_\text{L,R}$ is made explicit.

To avoid this complication, we compute the density matrix by instead evaluating
the expectation values of a tomographically complete set of observables, built
using the edge mode operators defined in Eq.\,\eqref{e:qb}. 
With these, we reconstruct the density matrix on the two qubit system residing on the edge using
\begin{equation}
\label{eq:tomo}
\rho(t) = \frac{1}{4}\sum_{\substack{i,j \in \\ \{0,x,y,z\}}} \langle \mathcal{O}^i_L \mathcal{O}^j_R (t)\rangle \sigma_L^i \otimes \sigma_R^j\,,
\end{equation}
where a superscript of 0 indicates that we take an identity operator in the
relevant Hilbert space. Beyond providing a computational advantage in the
following analysis, this method of evaluating the density matrix also provides
an intuitive picture for how one would characterize a memory experimentally.
The experimenter would measure a tomographically complete set of observables
until sufficient statistics are obtained to accurately determine the
expectation values in Eq.\,\eqref{eq:tomo} and hence reconstruct the density 
matrix \cite{Hradil:1997tt,nielsen}. This procedure, known as \emph{quantum state tomography}, has been implemented in a wide variety of 
quantum information experiments \cite{White:1999kd,Roos:2004hm,Steffen:2006cj,Lvovsky:2009uw}.
In general, at least $d^2-1$ different measurement settings are required to fully characterize a $d$-dimensional state. 

With the edge state determined, only the computational challenge of obtaining the effective channel $\mathcal{E}_t$ remains. Since $\mathcal{E}_t$ is linear, we can always think of it as acting on a vector space spanned by the Hilbert-Schmidt scalar products of basis elements, usually written as $\ketbra{m}{n}$, and find its matrix elements in that basis. In our case, the matrix form can be obtained from the numerical values of $\mathcal{E}_t(\ketbra{m}{n})$, using a Z-spin basis, letting
$\ket{m},\ket{n} \in
\{\ket{\uparrow ,\uparrow},\ket{\downarrow ,\downarrow},\ket{\uparrow,
\downarrow},\ket{\downarrow ,\uparrow}\}$. However, our procedure in Eq.\,\eqref{eq:tomo} 
only allows us to evaluate the map on elements of the form $\ketbra{m}{m}$, and so producing all the operator basis elements seems out of reach. Thankfully, there exists an algorithm for producing the off-diagonal elements, namely
\begin{align}
  &\mathcal{E}_t(\ketbra{m}{n}) \\
  &= \mathcal{E}_t(\ketbra{+}{+})+ i\mathcal{E}_t(\ketbra{-}{-})
  - \frac{1+i}{2} \mathcal{E}_t(\ketbra{n}{n})\nonumber \\
  &- \frac{1+i}{2}
  \mathcal{E}_t(\ketbra{m}{m})\,,\nonumber
\end{align}
where $\ket{+} = 1/\sqrt{2} ( \ket{m}+\ket{n} )$ and $\ket{-} = 1/\sqrt{2} ( \ket{m}+i\ket{n})$. This entire process of characterizing a map by performing state tomography on 
the output of a set of judiciously chosen inputs is often referred to as quantum \emph{process} 
tomography \cite{nielsen}.
Performing this calculation at every time step allows us to track the progressive degradation of the memory and calculate any of the single-shot 
quantities defined in Section \ref{ssec:QMem}. We do so for a number of disorder configurations 
and interaction strengths, obtaining a comprehensive picture of the edge's response to both.

\section{Results}

In this section, we present the numerical results obtained for 
the disordered SPT model with Ising-type perturbations defined in
Eq.\,\eqref{eq:H}. 
We carried out simulations for a system size $N =14$ and $100$ disorder realizations.
The Hamiltonian parameters used are $J \in \{0.1,0.075,0.05,0.025\}$ and
$\Delta \in \{0.5,1.0,1.5,2.0,2.5,3.0,3.5\}$, where we
deliberately chose some values $\Delta\geq2$ to study the regime where disorder competes with the bulk gap. 

All shown time traces were simulated from $t = 0$ to $2000$ (in units of inverse
energy of the $XZX$ coupling) by integrating Schr\"odinger's equation with a
step size of $\Delta t = 0.1$. For each disorder realization, we use
20 different initial states to obtain full tomographic information
of the two-qubit subspace.
At every integer time
slice, we calculate the expectation value of all 15 combinations of the
unperturbed edge mode operators given in Eq.\,\eqref{e:qb}.
Tracking these quantities over time allows us to 
reconstruct the full channel that represents the time evolution
reduced to the edges as laid out in Eq.\,\eqref{e:chan}.

\begin{figure}
\includegraphics[width=0.45\textwidth]{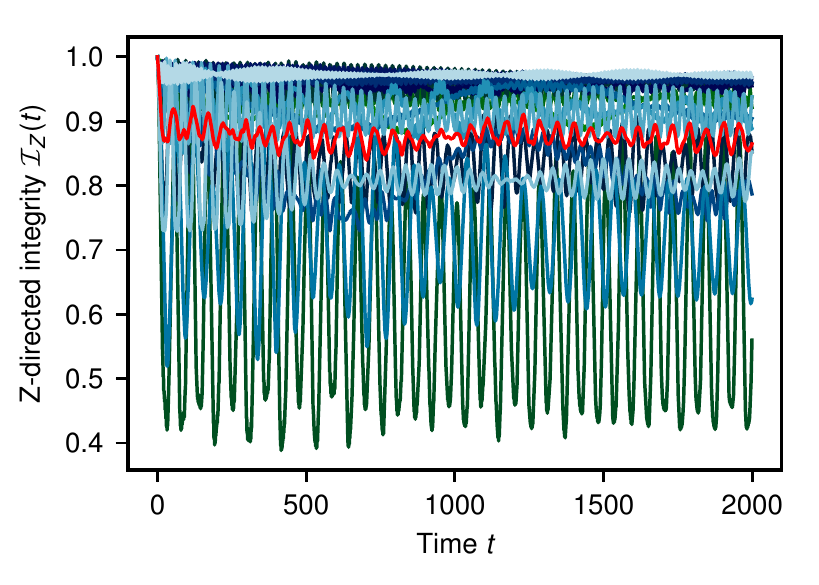}
  \caption{
    This plot shows the $Z$-directed integrity ${\cal I}_Z(t)$ as a function
    of time for 15 random samples (blue/green shades) and their average (red).
    The Hamiltonian parameters in this plot are $J=0.075$ and $\Delta=1.0$.
  The system is
  of size $N=14$ and we simulated the time evolution until $t=2000$ using step
  sizes of $\Delta t = 0.1$.
}
\label{f:sample}
\end{figure}

As discussed in Section~\ref{ssec:QMem}, we will first investigate directed
integrities which, individually, are only a measure of the quality of a
classical memory. When all three Pauli directions are taken together, however,
they serve as an upper bound for the integrity of the quantum channel.
We find that the behaviour of each instance varies considerably between 
different disorder realizations. To demonstrate this issue,
in Fig.\,\ref{f:sample} we show the time traces of the $Z$-directed
integrity
${\cal I}_Z(t)$ for 15 randomly chosen instances of disorder with Hamiltonian
parameters $J=0.1$ and $\Delta = 1.5$. The red curve is an
average over the 15 traces, the traces themselves are shown in shades of green
and blue. 

Though this average looks well-behaved, the obtained traces show a wide variety of behaviour from one realization to the next. 
It is useful to split the traces into two groups:
those which decreases exponentially with gentle fluctuations, and those which oscillate wildly within an exponentially decaying envelope. The former of these should dominate any average, and so we expect that our disorder-averaged quantities should softly decay, a finding consistent with previous results 
in related work \cite{bahri2015localization}. 

\begin{figure*}
\includegraphics[width=0.9\textwidth]{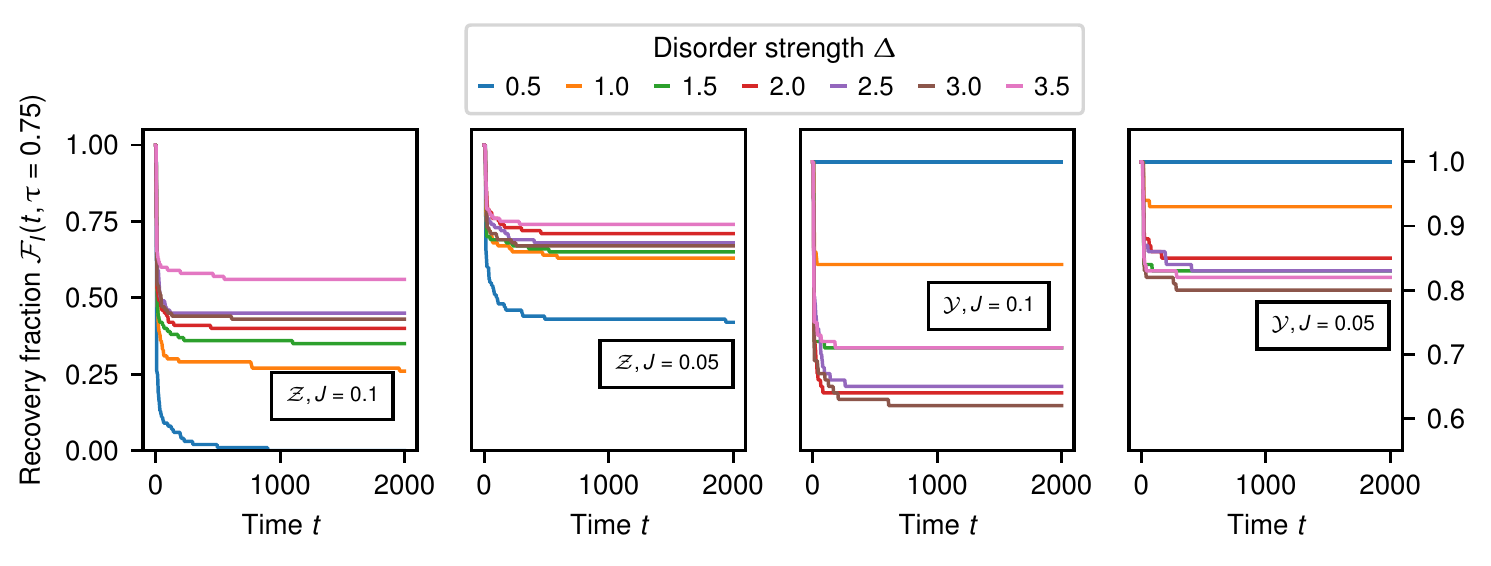}
  \caption{
  Recovery fractions of directed integrities for a fixed threshold as a function of time. 
  The left two panels show data for two different interaction
  strengths $J$ for the $Z$-directed integrity $\mathcal{F}_{{\cal I}_Z}(t,\tau=0.7)$.
  The panels on the right show the
  same interaction strengths $J$ for the $Y$-directed integrity $\mathcal{F}_{I_Y}(t,\tau=0.7)$. 
  Note the different scale of the y-axes.
  Colours encode the disorder strength $\Delta$. 
  The system is
  of size $N=14$ and we simulated the time evolution until $t=2000$ using step
  sizes of $\Delta t = 0.1$.
}
\label{f:pvst}
\end{figure*}

While this would usually be the end of a story featuring disorder, the appearance of rapidly fluctuating traces is very concerning. Since a low value in the integrity and coherent information both indicate the formation of entanglement, we can only conclude that the edge state is oscillating between states of high and low bulk entanglement.
This scenario constitutes a fiendish obstacle for any experimental
implementation. Even if one could in principle decode the stored data by 
picking a moment of low entanglement, the experimenter would need detailed knowledge about
the system dynamics to do so reliably. One might hope that these wild fluctuations are limited to this single integrity plot, indicating a pathology with limited scope. But, as we shall see, they appear in a sizeable fraction of disorder realizations and so are intrinsic to the dynamics of this system.

Given the extreme variation between disorder realizations, we find ourselves in a circumstance where disorder averaging would impede our ability to determine the suitability of this system to act as a quantum memory. Instead, we will define a phenomenological quantity appropriate
for the given problem, which we call the recovery fraction 
\begin{equation}
  \fr_A(t,\tau) := \frac{N_\text{rec}}{N_\text{tot}} \,,
\end{equation}
where $N_\text{rec}$ is defined as the number of instances for which the 
measured quantity $A$ is larger than $\tau$, i.e. 
$|\{A(t) | A(t')>\tau \quad \forall t'\leq t\}|$. The coefficient $\tau$ captures a 
minimum threshold that a hypothetical experimenter would need in order to 
recover the encoded information. In the limit of infinite sample size, 
the recovery fraction converges to a probability. This recovery probability 
gives the relative proportion of systems that \emph{reliably}, meaning for
all times smaller than $t$, stay above the recovery threshold $\tau$.
\begin{figure*}
\includegraphics[width=0.9\textwidth]{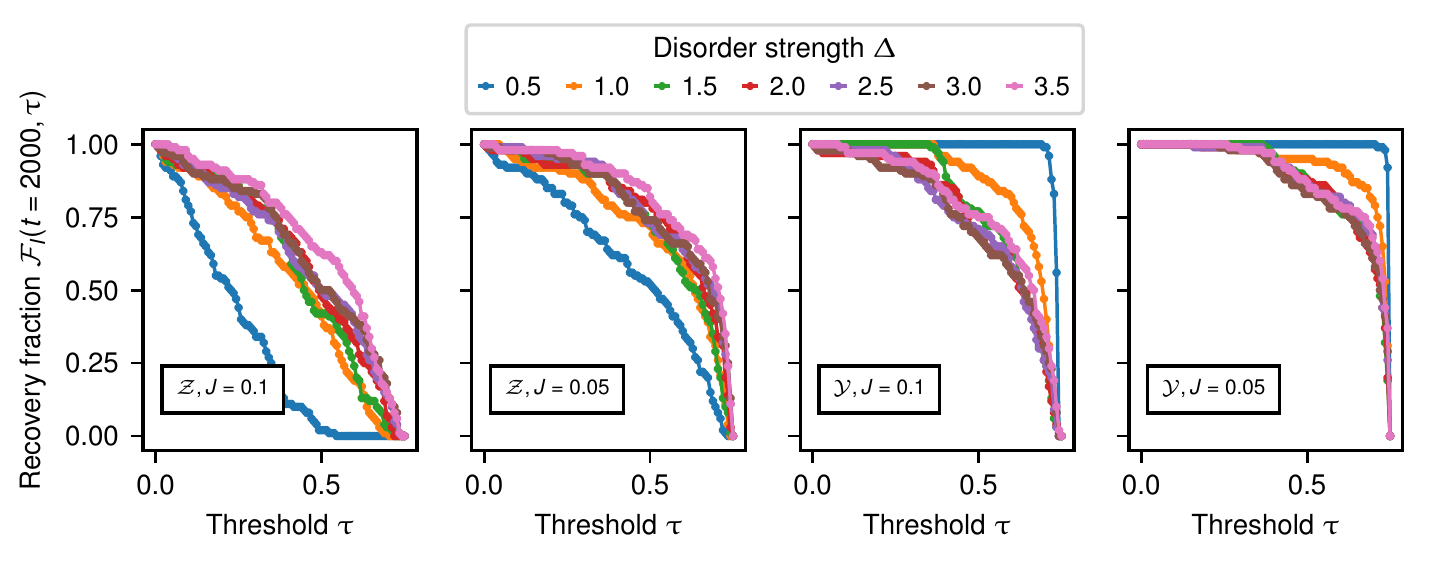}
  \caption{
  Final recovery fraction of directed integrities at $t=2000$ 
  as a function of the threshold $\tau$.
  The panels on the left show data for two different interaction
  strengths $J$ for the $Z$-directed integrity $\mathcal{F}_{{\cal I}_Z}(t=2000,\tau)$. 
  The panels on the right show the
  same interaction strengths $J$ for the $Y$-directed integrity $\mathcal{F}_{I_Y}(t=2000,\tau)$. 
  Colours encode the disorder strength $\Delta$. 
  The system is of size $N=14$.
}
\label{f:icutoff}
\end{figure*}

\subsection{Directed integrities}

We first investigate the directed integrities ${\cal I}_X(t),I_Y(t),{\cal I}_Z(t)$, which quantify how well the two initially orthogonal states in the $X,Y,$ or $Z$ logical basis remain so. Each of these gives an upper bound to the true integrity, but there are significant differences between these bases which require comment.

As described above, we will look at the recovery fraction to identify which
realizations would be unusable due to rapid oscillatory behaviour.
Fig.\,\ref{f:pvst} shows
the time evolution of the recovery fraction for each directed integrity. 
The left panels show the recovery fraction for the $Z$-directed integrity
$\fr_{{\cal I}_Z}(t,\tau=0.7)$ for interaction strengths $J \in \{0.1,0.05\}$.
Colours encode the different disorder strengths. For any
combination of Hamiltonian parameters, we find a steep initial
drop of the recovery probability followed by a long plateau.
When tuning the strengths of the interactions and disorder, we find
that decreasing interactions and increasing disorder increases
the apparent saturation value of the recovery probability. This is plausibly explained by the sharper localization behaviour in such conditions, a mechanism that
conceivably improves the storage of classical information.
The sharp initial drop-off is both intriguing and highly disturbing, as it implies 
that even for relatively weak interaction strengths (e.\,g.\,$J = 0.05$), a 
huge fraction of disorder realizations fall below threshold immediately. 
The loss of 25\% to 50\% of these shows that, while the overall decay profile may be exponential, the fine structure of the evolution would preclude us from saying that the information is protected in any practical sense. We would like to point out that the $X$-directed integrity shows the same qualitative behaviour and thus is not displayed separately here.

When analysing the recovery fraction for the $Y$-directed integrity 
$\fr_{I_Y}(t,\tau=0.7)$ shown in the two panels 
on the right, we find a sizeable quantitative increase of the recovery fraction
accompanied by a similar qualitative effect of
interactions, namely the recovery probability decreases with interaction strength. However, its disorder dependence differs significantly
from the ${\cal I}_Z(t)$ case. Here, increasing the disorder strength \emph{decreases} the integrity of the $Y$-basis.   
Moreover, the recovery fraction is the lowest when the disorder
strength is large enough to cancel the mean bulk gap ($\Delta \geq 2$). While initially shocking, the large 
quantitative and qualitative difference between bases actually mirrors results from previous 
numerical studies performed by some of the authors \cite{ourSPT}, where the perturbed edge zero 
mode operators showed a similar disorder and interaction dependence.

While the time dependence of $\fr_I$ reveals curious behaviour, in practice, an experimenter will be interested in how well the bit
can be recovered after time evolution. It is thus instructive
to consider the final value of our simulation at $t=2000$ and 
plot this value as a function of the threshold $\tau$, ensuring that we 
have not artificially depressed the recovery fraction by choosing an overly optimistic threshold. The
resulting plots are shown in Fig.\,\ref{f:icutoff}. Again, the
colour encodes disorder and the two panels on the left show two
interaction strengths $J \in \{0.1,0.05\}$ for the
final recovery fraction of the $Z$-directed integrity
$\mathcal{F}_{{\cal I}_Z}(t=2000,\tau)$. Here, we find that the recovery fraction starts at one and smoothly goes to zero when the threshold reaches one. 
Again, the interaction plays a crucial role here, depressing the recovery fraction uniformly 
as it increases. The disorder blunts this effect as before, but the effect saturates when it is of the order of the mean bulk gap. 
When turning to the final recovery fraction of the $Y$-directed integrity 
$\mathcal{F}_{I_Y}(t=2000,\tau)$ shown in the two panels 
on the right, we again see an extreme sensitivity to the disorder strength, 
with a final recovery fraction which is almost identically 1 for all but the most aggressive values of $\tau$ while disorder is very low ($< 0.5$), with an immediate drop-off for higher disorder strengths. What becomes more apparent is disorder's homogenization of the bases, improving the integrity of the $X$ and $Z$ encodings while simultaneously degrading the $Y$ encoding so that the final recovery fractions are similar.
\begin{figure}[H]
\includegraphics[width=0.45\textwidth]{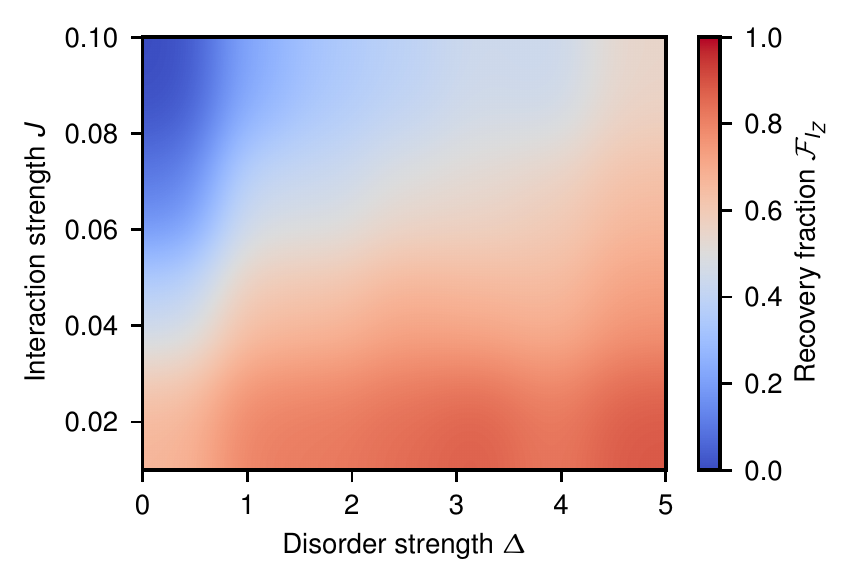}
  \caption{In this heat map, we show the disorder and interaction dependence of
  the final recovery fraction $\mathcal{F}_{{\cal I}_Z}(t=2000,\tau=0.7)$ for a fixed
  threshold for the $Z$-directed integrity.
}
\label{f:Zhm}
\end{figure}
\begin{figure}[H]
\includegraphics[width=0.45\textwidth]{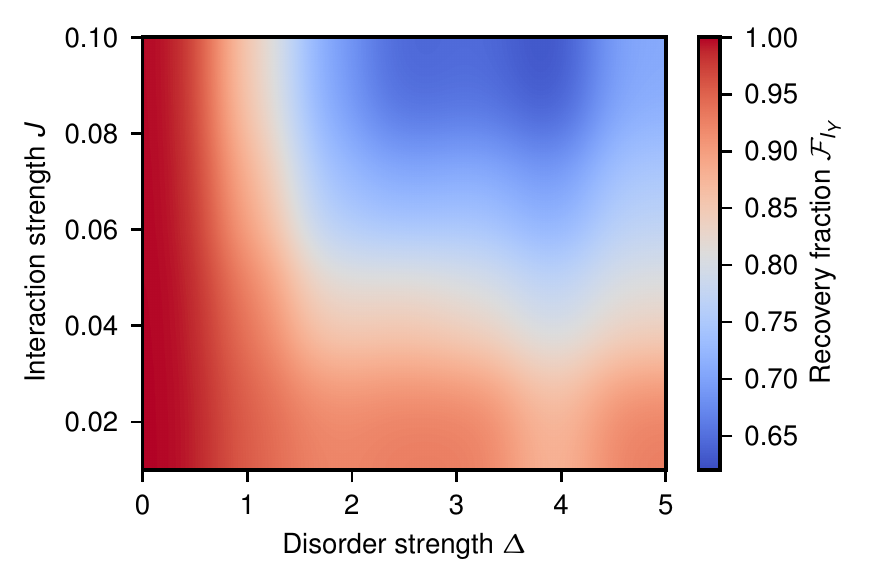}
  \caption{In this heat map, we show the disorder and interaction dependence of
  the final recovery fraction $\mathcal{F}_{I_Y}(t=2000,\tau=0.7)$ for a fixed
  threshold for the $Y$-directed integrity.
}
\label{f:Yhm}
\end{figure}

\begin{figure*}
\includegraphics[width=0.9\textwidth]{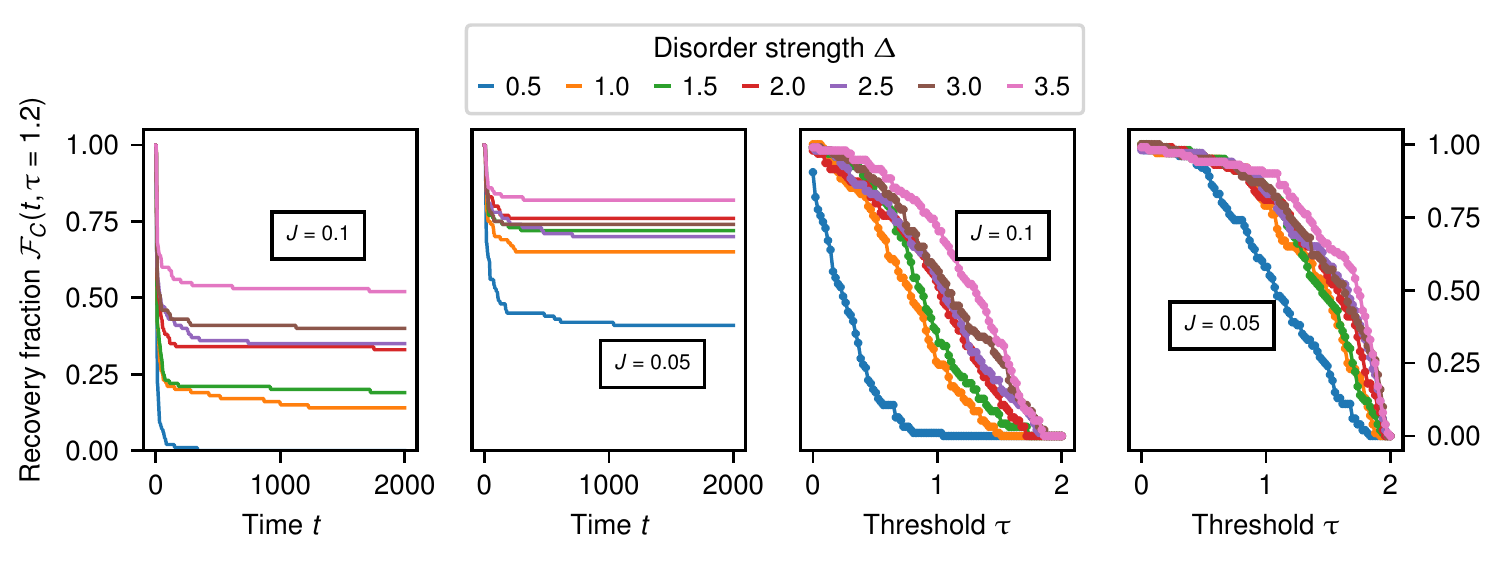}
  \caption{
  This plot shows data for the coherent information for interaction strengths
  $J \in \{0.1,0.05\}$. Panels on the left show the time evolution
  of the recovery fraction of the coherent information
  $\mathcal{F}_{C}(t,\tau=1.2)$.
  Colours encode the disorder strength $\Delta$. The system is of size $N=14$.
  Simulations run up to $2000$ tunnelling times with a step size of $\Delta t = 0.1$.
  Panels on the right show the final recovery fraction of the coherent
  information $\mathcal{F}_{C}(t=2000,\tau)$ 
  as a function of the threshold $\tau$ for the same interaction and disorder 
  strengths.
}
\label{f:ci}
\end{figure*}

These results are nicely summarised in Fig.\,\ref{f:Zhm} and \,\ref{f:Yhm}, heat maps of $\mathcal{F}_{{\cal I}_Z}(t,\tau=0.7)$ and $\mathcal{F}_{I_Y}(t,\tau=0.7)$ respectively. Here we can clearly see the monotonic improvement of the $Z$-directed integrity with increasing disorder and decreasing interaction strength, while the $Y$-directed integrity shows a clear dip as $\Delta$ approaches 2. We do see a resurgence in both plots when $\Delta$ moves beyond 4, but this corresponds to the disorder scale becoming the dominant energy scale in the problem, playing a similar role to the mean gap in the original problem.

Taken together, the directed integrities paint a curious picture. Classical bits encoded using any basis will be susceptible to degradation from interactions. Bizarrely, our ability to attenuate this with disorder depends more on the chosen edge basis than the interaction strength, with the $Y$-basis being clearly superior for weak disorder. 
We would like to stress again, that considering single directed integrities
only allows conclusions about the classical information storage. 
The actual integrity is the minimum of all possible encodings, and is thus upper bounded by the worst of our directed integrity results, e.g. the $Z$-directed integrity. Since this is improved by introducing disorder, this suggests that disorder does in fact help its ability to store quantum information, albeit at the cost of reducing its classical information storage capabilities, which are captured by the $Y$-directed integrity. To confirm this, we will now calculate the coherent information of the dynamical map.

\begin{figure}
\includegraphics[width=0.45\textwidth]{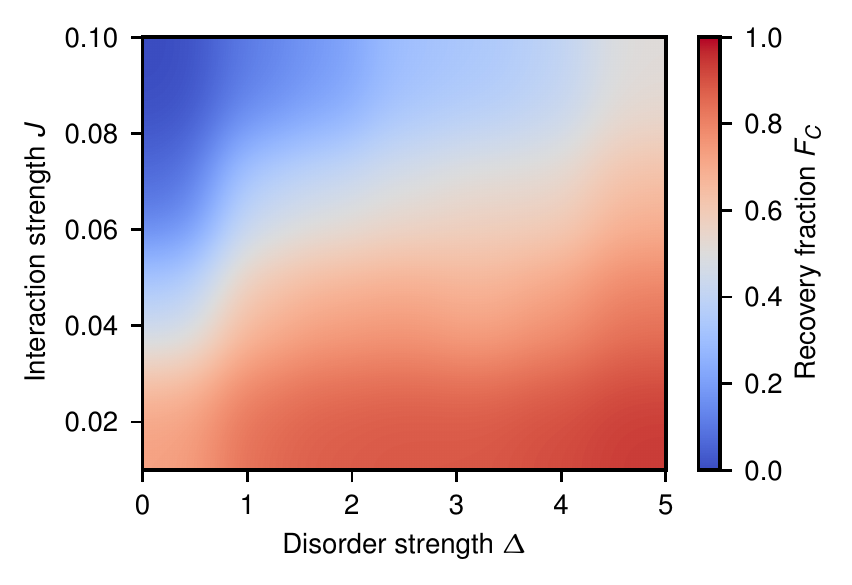}
  \caption{In this heat map, we show the disorder and interaction dependence of
  the final recovery fraction of the coherent information
  $\mathcal{F}_{C}(t=2000,\tau=1.2)$.
}
\label{f:Chm}
\end{figure}

\subsection{Coherent information}

In this part, we carry out the very same analysis as for the directed
integrities but now we use the coherent information $C(t)$ which gives a lower bound
to the single-shot capacity $Q_1(t)$ of our system. $Q_1(t)$ quantifies the number of qubits which could be reliably be extracted from our system after our time evolution using some optimal single-shot encoding strategy. The coherent information then corresponds to fixing an encoding strategy, and thus both range from zero to two.

The time dependence
of the recovery fraction of the coherent information
$\mathcal{F}_C(t,\tau=1.2)$ for a fixed threshold and different interaction
strengths is shown in the left panels of Fig.\,\ref{f:ci}. The behaviour shown
is strikingly similar to the $Z$-directed integrity, showing the same immediate
decline in recovery fraction, but now for a quantity which is genuinely quantum. 
Decreasing the interaction strength and increasing the disorder strength 
both improve the coherent information monotonically, a feature also shared with the $Z$-directed integrity. 
 These similarities continue in the final recovery probability of the coherent information $\mathcal{F}_C(t=2000,\tau)$ as a function of cutoff $\tau$ shown in the panels on the right in Fig.\,\ref{f:ci}, showing that most of the improvement comes from reducing the interaction strength.

If we now again consider the final recovery probability of the coherent information
$\mathcal{F}_C(t=2000,\tau=1.2)$ with fixed threshold as a function of disorder
and interaction strength shown in Fig.\,\ref{f:Chm}, 
we again see that high disorder and low interaction
lead to the highest final recovery fraction of the coherent information.
The figure looks very similar to what we have obtained for the $Z$-directed
integrities, confirming that the ability of these to store information is the limiting factor in edge's ability to preserve quantum information reliably.

\section{Conclusions}

Using edge modes of topologically protected systems is a promising avenue
for quantum information tasks. It is, however,
largely unclear how stable information encoded into these systems
will actually be when the system is perturbed away from its solvable point by spurious interactions and disorder, both of which are omnipresent in realistic
experimental realizations. Here, we present numerical results
for the dynamical recoverability of information -- both classical and quantum --
encoded into the edge mode subspace of an
$XZX$ cluster Hamiltonian with both tunable many-body interactions and disorder
as a perturbation. We reconstruct the edge state
density matrix to obtain the channel which represents the time evolution.
With these data, we are in position to calculate measures of classical
information, namely directed integrities; and quantum information, namely the
coherent information.

Studying the directed integrities, we find that not only disorder strength 
and interaction strength change the amount of recoverable information, but also that
the specific encoding direction plays a crucial role in determining the coherence of the encoded bits. The best encoding is found to be in the
$Y$ basis of the unperturbed Hamiltonian, yielding very stable bits for a 
large range of interaction strengths and low disorder. The narrative of disorder
aiding the encoding is wrong for this setting. It holds, however, for an encoding
in the $Z$ or $X$ bases, where the detrimental effect of interaction strength can
be counteracted by introducing disorder.

For quantum information, we recover and provide further substance to 
the narrative that disorder increases the ability of the system to store quantum information. Interactions
decrease the quality of the memory in a similar way to the classical encoding
in $Z$ or $ X$ direction.

To distinguish well- and ill-behaved disorder realizations, we set a minimum performance threshold and evaluated the fraction of runs which stayed above it over the course of the evolution. Quite surprisingly, this fraction failed to decay smoothly over time, as one would expect if there was gradual decoherence of the edge states. Instead, the recover fraction dropped precipitously, indicating that a large fraction of disorder realizations describe systems that would be worthless for information storage purposes. Furthermore, these may not have an appreciable effect on disorder averages, as they are rapidly oscillatory instead of simply dropping to zero immediately.
It seems fair to say that the interplay of disorder,
interactions and SPT order is less understood than previously anticipated. Our
findings are compatible with those in Ref.\,\cite{ourSPT}, where a similarly
multi-faceted picture for the edge mode support was found using a completely
different method. These results demand more foundational work possibly
including an extension of perturbation theory to systems with topologically
protected degeneracies.

Our work covers information loss due to unitary dynamics under perturbed
Hamiltonians, but real world experiments are also coupled to the outside
world leading to further decoherence. Since localization effects
are also expected to be diminished in the
presence of dissipation \cite{vanNieuwenburg2017,Luschen2017}, 
the amount of information protection induced by
disorder seems questionable. Given that open system dynamics are also capable of corrupting topological data directly, even without the presence of many-body interactions \cite{2012majorana_decoherence, 2012majorana_poison, 2013majorana_recovery}, investigating the circumstances where disorder, interactions, and a particle reservoir are all present is likely to yield fascinating results.
\vspace{4pt}
\section{\co-emission table}

Here, we summarize the climate expenses of the simulations presented in this
work.

\begin{figure}[H]
\begin{center}
\begin{tabular}[b]{l c}
\hline
\textbf{Numerical simulations} & \\
Total Kernel Hours [$\mathrm{h}$]& 308000\\
Thermal Design Power per Kernel [$\mathrm{W}$]& 5.75\\
Total Energy Consumption of Simulations [$\mathrm{kWh}$] & 1771\\
Average Emission of CO$_2$ in Germany [$\mathrm{kg/kWh}$]& 0.56\\
Total \co-Emission from Numerical Simulations [$\mathrm{kg}$] & 992\\
Were the Emissions Offset? & \textbf{Yes}\\
\textbf{Transportation} & \\
Total \co-Emission from Transportation [$\mathrm{kg}$] & 0\\
Were the Emissions Offset? & \textbf{Yes}\\
\hline
Total \co-Emission [$\mathrm{kg}$] & 992\\
\hline
\end{tabular}
\caption{
  {Estimated climate footprint of the computations
  presented in this paper. Prototyping is not included in these calculations.
  Estimations have been calculated
  using the examples of Scientific CO$_2$nduct \cite{scicon2019} and are correct
  to the best of our knowledge.}}
\end{center}
\end{figure}
\section{Acknowledgements}

We would like to thank Ingo Roth for discussions on optimization schemes. NW acknowledges funding support from the European Unions Horizon 2020 research and innovation programme under the Marie Sklodowska-Curie grant agreement No.750905. This work has been supported by the ERC (TAQ), the DFG (CRC 183, FOR 2724, EI 519/7-1), and the Templeton Foundation. This work has also received funding from the European Union’s Horizon2020 research and innovation programme under grant agreement No 817482 (PASQuanS).

\bibliography{SPT_bibliography}

\end{document}